\documentstyle[12pt,aaspp4]{article}

\begin{document}

\title{Spectroscopy of Giant Stars in the Pyxis Globular Cluster}

\author{Christopher Palma}
\affil{Department of Astronomy, University of Virginia \\
       Email:  cp4v@virginia.edu}
\authoraddr{P. O. Box 3818, Charlottesville, VA  22903-0818}

\author{William E. Kunkel\altaffilmark{1}}
\affil{The Observatories of the Carnegie Institute of Washington \\
       Email:  kunkel@jeito.lco.cl}
\authoraddr{813 Santa Barbara Street, Pasadena, CA 91101}
\altaffiltext{1}{Las Campanas Observatory, Casilla 601, La Serena, Chile}

\and

\author{Steven R. Majewski\altaffilmark{2}}
\affil{Department of Astronomy, University of Virginia \\
       Email:  srm4n@virginia.edu}
\authoraddr{P. O. Box 3818, Charlottesville, VA  22903-0818}
\altaffiltext{2}{David and Lucile Packard Foundation Fellow; Cottrell Scholar
of the Research Corporation; National Science Foundation CAREER Fellow;
Visiting Associate, The Observatories of the Carnegie Institution of Washington}

\begin{abstract}

The Pyxis globular cluster is a recently discovered globular cluster
that lies in the outer halo ($R_{gc} \sim 40$ kpc) of the Milky Way.
Pyxis lies along one of the proposed orbital planes of the Large
Magellanic Cloud (LMC), and it has been proposed to be a detached LMC
globular cluster captured by the Milky Way.  We present the first
measurement of the radial velocity of the Pyxis globular cluster based
on spectra of six Pyxis giant stars.  The mean heliocentric radial
velocity is $\sim\,$36 km/sec, and the corresponding velocity of Pyxis
with respect to a stationary observer at the position of the Sun is
$\sim-191$ km/sec.  This radial velocity is a large enough fraction of
the cluster's expected total space velocity, assuming that it is bound
to the Milky Way, that it allows strict limits to be placed on the
range of permissible transverse velocities that Pyxis could have in the
case that it still shares or nearly shares an orbital pole with the
LMC.  We can rule out that Pyxis is on a near circular orbit if it is
Magellanic debris, but we cannot rule out an eccentric orbit associated with
the LMC.  We have calculated the range of allowed proper motions for
the Pyxis globular cluster that result in the cluster having an orbital
pole within $15^{\circ}$ of the present orbital pole of the LMC and
that are consistent with our measured radial velocity, but verification
of the tidal capture hypothesis must await proper motion measurement
from the Space Interferometry Mission or HST.  A spectroscopic
metallicity estimate of $[$Fe/H$] = -1.4\pm0.1$ is determined for Pyxis
from several spectra of its brightest giant; this is consistent with
photometric determinations of the cluster metallicity from isochrone
fitting.

\end{abstract}


\section{Introduction}

Evidence continues to accumulate that the outermost Milky Way globular
clusters may not have originated in the same process that formed the
inner globular clusters.  Based on the recognition that the second
parameter effect of horizontal branch morphology in globular clusters
is found predominantly among outer halo ($R_{gc} > 8$ kpc) clusters,
\cite{SZ} proposed that the outermost globular clusters may have formed
in chemically distinct ``fragments'' that later fell into the Milky Way
halo.  Building on the suggestion by \cite{kd76} that several red
horizontal branch (second parameter) globular clusters were potentially
associated with the Magellanic Plane group of dwarf galaxies,
\cite{srm94} showed that there is a planar alignment between a
particular sample of second parameter globular clusters and the Milky
Way dwarfs.  Recently, \cite{palma99} reaffirmed that there may be a
dynamical relationship between the second parameter globular clusters
and the Milky Way dwarf satellites.

The Pyxis globular cluster (\cite{idk95}; \cite{d95}) at $R_{gc} = 41$
kpc (Sarajedini \& Geisler 1996) defines the inner edge of the
prominent gap in the globular cluster radial distribution between $40
\lesssim R_{gc} \lesssim 60$ kpc.  The presence of this gap has been
used to argue that the primordial Galactic globular cluster system ends
at $\sim\,$40 kpc while the distant, $R_{gc} > 60$ kpc clusters
originated in Galactic satellite dwarf galaxies (e.g., \cite{zinn85}).
Although Pyxis lies among the ``inner group'' of globular clusters
(i.e., inside the gap in $R_{gc}$), Irwin et al.\ (1995) propose that
Pyxis may be a captured LMC globular cluster based on the young age
they infer for the globular cluster and on its proximity to the plane
of the LMC orbit derived from the \cite{jones} proper motion.  Further
support for the tidal capture hypothesis comes from \cite{palma99},
where a statistical analysis of the likely orbital poles of the
Galactic satellite galaxies and the globular clusters identifies Pyxis,
NGC 6229, NGC 7006, and Pal 4 as the globular clusters most likely to
share a common orbital pole with either the Magellanic Plane galaxies
(the LMC, the SMC, Draco, and Ursa Minor) or the Fornax--Leo--Sculptor
Stream galaxies.  However, these postulations on the origin of Pyxis
have been made without the benefit of any kinematical data on the
cluster.

Although deep photometry of Pyxis exists (Sarajedini \& Geisler 1996),
no spectroscopic observations have been published.  Indeed, Pyxis is
one of the last few known globular clusters lacking a radial velocity
(cf.\ Harris 1996).  We report here on du~Pont 2.5-m Telescope
spectroscopic observations of Pyxis stars (\S 2).  With our derived
radial velocity for the cluster, we re-address the stripped LMC
hypothesis for Pyxis' origin (\S 3), but point out that, in the end, we
can only make predictions on the proper motions expected under this
scenario.  Unfortunately, the proper motion is required for a
definitive solution to the question of the cluster's origin.

\section{Observations}

On the nights of 17 January and 20 January 1997, the 2.5-m du Pont
Telescope at the Las Campanas Observatory was used to obtain spectra of
Pyxis giant stars with magnitudes of $R \sim 18$.  A finding chart for
these stars made from the Digitized Sky Survey (\cite{dss}) is
presented as Figure 1.  Observations of the Pyxis stars made use of the
modular spectrograph and the SITe2 detector with a 1200 lines/mm
grating.  The selected wavelength region with this setup was
approximately 7700 -- 8750 \AA\ at $\sim\,$1.3 \AA\ per pixel
resolution.  Typical exposure times were 900 seconds per observation,
which provided enough signal-to-noise to measure radial velocities.

Radial velocities were measured for six Pyxis giant stars by
cross-correlating their spectra against those of bright radial velocity
standard stars.  The cross-correlation peak yielded the adopted radial
velocity difference between the Pyxis stars and the standard.  The star
HD80170, a K5 giant, was observed multiple times on both nights to
serve as the main standard.  Two other HD stars and 11 LMC globular
cluster stars were also observed as radial velocity calibration
objects.  All standards with known radial velocities were used to
remove any nightly zero point offsets in the radial velocities
determined by cross-correlating against the HD80170 template.  The
Pyxis stars were observed multiple times, and the radial velocity for
each star was determined by taking the median of the values for a
particular star.  In Table 1 we present the positions, magnitudes, and
velocities of the stars observed.  We find a  mean radial velocity for
the six stars of $34.3$ km/sec with $\sigma = 4.6$ km/sec.  If we
ignore the one outlier, Pyxis D, the mean radial velocity for the five
remaining stars increases slightly, to $35.9$ km/sec with $\sigma =
2.5$ km/sec.

\subsection{\ion{Ca}{2} Triplet Metallicity Estimate}

The wavelength range of our spectra includes the \ion{Ca}{2} triplet
lines at $\lambda\lambda$ 8498, 8542, and 8662 \AA.  \cite{ra97} measured
\ion{Ca}{2} equivalent widths for a large sample of globular cluster
stars and in a companion paper (Rutledge et al.\ 1997b) derived a
conversion from \ion{Ca}{2} reduced equivalent width, $W^{\prime}$, to
metallicity.  To determine a spectroscopic metallicity for Pyxis, we
attempted to measure the strengths of these lines in our spectra.  Of
the six Pyxis stars observed, only the brightest, Pyxis A, produced
equivalent width measurements with errors small enough to give a
reliable estimate of the metallicity.  
The technique used to determine
the equivalent widths for the Pyxis A \ion{Ca}{2} lines was nearly
identical to that used in \cite{ra97}.

For each observation of Pyxis A, the continuum was determined by
linearly interpolating the average intensity in the \cite{ra97}
continuum bandpasses.  The equivalent width was then calculated by
integrating the difference between the fit continuum and the line
feature over the line bandpass.  The line feature was fit with a
Gaussian function, and the integral was performed using the fit rather
than numerically integrating the data since \cite{ra97} concluded that
this technique is preferable to direct numerical integration.  The
three individual lines were combined into a single index, $\Sigma$Ca,
following the method of \cite{ra97}, where $\Sigma$Ca $=
0.5\lambda_{8498} + \lambda_{8542} + 0.6\lambda_{8662}$.  For Pyxis A,
the result is $\Sigma$Ca $= 4.5\pm0.2$ \AA.

The \cite{ra97} method for converting $\Sigma$Ca to metallicity
requires the reduction of the equivalent width to the value for giants
at the level of the horizontal branch, $W^{\prime}$, so that a mean
value for all cluster stars can be obtained.  This is done by adopting
a slope $\Delta(\Sigma{\rm Ca})/\Delta(V_{HB} - V)$ and extrapolating
the calculated width to the expected value at the magnitude of the horizontal
branch, $V_{HB}$.  All of the
published photometry for Pyxis stars has used the $B$ and $R$ bands, so
we have had to approximate $\Delta(V_{HB} - V)$ using color-color
relations for giant stars.  Adopting the photometry from \cite{sg96}, we find
that Pyxis A has $R = 17.08$ and that the horizontal branch of Pyxis is
at $R = 18.75$.  \cite{saao} find an almost linear relationship between
$(B-V)$ and $(V-R)$ for giant stars, so we have determined rough $V$
magnitudes for Pyxis A and the Pyxis horizontal branch stars by estimating a $(V-R)$
color from the $(B-R)$ color given in \cite{sg96}.  We estimate that
for Pyxis, $V_{HB} = 19.25$ and $V_{Pyxis A} = 17.77$, or
$\Delta(V_{HB} - V) = 1.48$.  In the table of globular cluster
properties by \cite{mwgc}, the magnitude of the horizontal branch is
also given as $V_{HB} = 19.25$, so the adopted $(V-R)$ colors are most
likely a good approximation to the true colors.

The reduced \ion{Ca}{2} equivalent width of Pyxis A is therefore
$W^{\prime} = 3.6$ if we follow \cite{ra97} and adopt a slope of
$\Delta(\Sigma{\rm Ca})/\Delta(V_{HB} - V) = 0.62$ \AA/magnitude.  This
value is simply an estimate, since there is a dispersion of 0.2 \AA\ in
the values of $\Sigma$Ca from the four individual observations and
since there is some uncertainty in $\Delta(V_{HB} - V)$, probably of
order 0.1 magnitudes.  However, we can use this determination to get an
estimate of the metallicity of Pyxis A for comparison with the
photometrically determined metallicity estimates of Irwin et
al.\ (1995) and \cite{sg96}.  Using the \cite{rb97} calibration of
$W^{\prime}$ to Zinn-West metallicity, the reduced equivalent width
measured for Pyxis A implies a metallicity of $[$Fe/H$]_{ZW} = -1.4 \pm
0.1$.  This value for Pyxis A is more metal-poor than the
photometrically derived values of \cite{sg96} and Irwin et al.\ (1995),
who estimate $-1.2\pm0.15$ and $-1.1\pm0.3$ respectively, however it is
consistent within the overlap of the $1\sigma$ error bars.  Any 
systematic error that leads to an underestimated equivalent width for Pyxis A
results in a smaller determined metallicity.  An error of 10\% in the
equivalent width measured for Pyxis A is enough to bring the
metallicity up to $-1.2$ and into better agreement with the photometric
values.  If the equivalent width measurement is correct, then it is
unlikely that the metallicity is much higher than $[$Fe/H$]_{ZW} =
-1.4$, since an unlikely error in the $V$ magnitude of Pyxis A of 0.7
magnitudes is required to raise the metallicity of the star to
$-1.2$.

\section{Discussion}

The observations presented here were partially motivated by the
possibility that the Pyxis globular cluster was captured from the LMC
by the Milky Way.  This assertion was originally made by Irwin et
al.\ (1995), who noted that Pyxis, at $(l,b) = (261.3,7.0)^{\circ}$, lies
within a few degrees of the orbital plane of the LMC determined from
the \cite{jones} proper motion.  Further support for this hypothesis is
provided by \cite{palma99}, who place Pyxis in a group with Pal 4, NGC
6229, and NGC 7006 as the most likely globular clusters to share a
common orbital pole with the Magellanic Plane galaxies (the LMC, the
SMC, Ursa Minor, and Draco).  Although full space motion information is
required to verify the Irwin et al.\ (1995) hypothesis, a radial
velocity can provide some constraints on the shapes of allowed orbits
for a cluster if the magnitude of the radial component is a significant
fraction of the expected magnitude of the space velocity.

The essence of the argument given in Palma et al.\ (2000) to support a
capture origin for the Pyxis globular cluster is as follows:  if it is
assumed that Pyxis was captured recently from the LMC by the Milky Way,
then the orbital pole of Pyxis is likely to be aligned with that of the
LMC (both the LMC and Pyxis are far enough from the Galactic Center
that precession will not significantly affect the positions of their
orbital poles over a Hubble time).  If one assumes rotation about the
Galactic Center, the direction of the orbital pole of the LMC can be
determined by taking the cross product of the Galactocentric radius
vector to the LMC and its space motion vector.  If one accounts for the
space velocity vector error, the position of the orbital pole of the
LMC can only be confined to a {\em family} of poles along an arc
segment in Galactocentric coordinates (cf.\ Figure 1 in Palma et
al.\ 2000).  Since the space motion of Pyxis is currently unknown, its
orbital pole is not well constrained.  However, the orbital pole can be
assumed to be perpendicular to its current Galactocentric position, so
the direction of Pyxis' orbital pole should lie on the great circle
that contains all possible normals to its current radius vector (cf.
\cite{lb295}).   Figure 2 shows an Aitoff projection of the sky in
Galactocentric coordinates.  The arc segment that defines the possible
locations of the LMC's orbital pole (based on the Jones et al.\ [1994]
proper motion, as adopted by Palma et al.\ [2000]) is shown as well as
the great circle along which lies all possible orbital poles of Pyxis.
That these two families of possible orbital poles for the LMC and for
Pyxis intersect (at Galactocentric $(l,b) = (163, -22)^{\circ}$)
indicates that it is possible for these two objects to share a coplanar
orbit with a common direction of angular momentum.  It now remains to
be seen whether our derived radial velocity can clarify whether this is
likely, i.e., is the orbital pole of Pyxis likely to be near the
crossing point of the two orbital pole families?

We start our analysis by adopting a simple strawman model wherein 
Pyxis is following a circular orbit that is
nearly coplanar with the orbit of the LMC. Simulations of tidal
stripping of dwarf galaxies by the Milky Way (Johnston 1998) show that
the debris is distributed around the orbit of the parent satellite with
a spread in energy given by 
\begin{equation}
                \Delta E= r_{\rm tide} {d\Phi \over dR} \approx
                \left( {m_{\rm sat} \over M_{\rm Gal}}\right)^{1/3} v_{\rm circ}^2
                \equiv f v_{\rm circ}^2
\end{equation}
where $r_{\rm tide}$ is the tidal radius of the satellite, $\Phi$ is
the parent galaxy gravitational potential, $v_{\rm circ}$ is the
circular velocity of the Galactic halo, $m_{\rm sat}$ is the
satellite's mass, $M_{\rm Gal}$ is the mass of the parent galaxy
enclosed within the satellite's orbit, and the last equality defines
the {\it tidal scale} $f$.  Thus, the spread in energy translates into
a characteristic angular width $f$ (in radians) to the debris.  Taking
reasonable values for the mass of the LMC and the Milky Way, the value
of $f$ for any debris pulled from the LMC corresponds to roughly
$15^{\circ}$.  In Figure 2, an arc segment along the great circle of
possible orbital poles for Pyxis is marked; this arc segment is defined
by a length within
$\pm15$ degrees of the intersection point with the possible orbital
pole of the LMC, and indicates expectations for the orbital poles of
LMC debris at the position of Pyxis on the sky. 

It is straightforward to derive the direction of the space motion
vector required for Pyxis to follow a circular orbit and have an
orbital pole along the arc segment in Figure 2, i.e.\ in the direction
of $\hat{P}$, or $(l,b) = (163,-22)^{\circ}$.  The geometry is
illustrated in Figure 3.  The Galactocentric, Cartesian radius vector
of Pyxis is $(X,Y,Z) = (-13.9,-38.6,4.8)$ kpc (assuming that the
distance from the Sun to the Galactic Center, $R_{0} = 8$ kpc, and
where $X_{\sun}\equiv-8.0$ kpc).  The unit vector $\hat{P}$ from the
Galactic center in the direction of the orbital pole at Galactocentric
$(l,b) = (163,-22)^{\circ}$ is $(X,Y,Z) = (-0.89,0.27,-0.37)$.  In
Figure 3, this vector has been translated to the location of Pyxis.
The vector that is mutually perpendicular to the Galactocentric radius
vector of Pyxis and to the orbital pole $\hat{P}$ gives the direction
of the space motion of the Pyxis globular cluster for a circular orbit
around the Galactic center.  The unit vector direction of this space
motion is $\hat{V}_{circ} = (V_{X},V_{Y},V_{Z}) = (-0.32,0.23,0.92)$.
Since our line of sight to Pyxis is mostly in the $- Y$ direction and
since $\vec{V}_{circ}$ is mostly in the $+ Z$ direction, clearly any
component of $\vec{V}_{circ}$ along our line of sight will be small.
Adopting the basic solar motion of $(9,12,6)$ km/sec (\cite{mb81}) and
a rotational velocity of the local standard of rest (LSR) of 220
km/sec, then the component of the Sun's velocity along the line of
sight from the Sun to Pyxis is $v_{\sun LOS} \sim-227$ km/sec (the
negative sign here indicates that the component of the Sun's velocity
with respect to the Galactic Standard of Rest along the line of sight
to Pyxis is in the sense of receding from Pyxis; see the Appendix for a
discussion of the sign conventions used in reducing the radial velocity
to a $v_{GSR}$).  Since the heliocentric radial velocity measured for
Pyxis is the difference between the intrinsic radial velocity of Pyxis
with respect to the position of the Sun, $v_{GSR}$, and the magnitude
of the Sun's velocity projected along the line of sight to Pyxis
($v_{helio} = v_{GSR} - v_{\sun LOS}$), the globular cluster would have
a large, positive heliocentric radial velocity if it were following a
circular orbit in the plane defined by the pole at $\hat{P}$.

It is the solar motion that dominates the radial velocity in the
circular orbit case.  For example, if we assume that Pyxis has a
velocity that is approximately the circular velocity of the Galaxy at
40 kpc, or $\sim\,$200 km/sec, its space motion would then be
$(V_{X},V_{Y},V_{Z}) = (-64,46,184)$ km/sec.   Since the Galactic
Cartesian unit vector in the direction of Pyxis from the Sun (the line
of sight) is $(X,Y,Z) = (-0.15,-0.98,0.12)$, the component of this
space velocity along the line of sight\footnote[3]{Again the negative
sign is in the radial system of the Sun, and it indicates that this
velocity points towards the Sun.  In the Galactic Cartesian system this
velocity is positive in the $X$, $Y$, and $Z$ directions.} would be
$(-64 \times -0.15)+(46 \times -0.98)+(184 \times 0.12)$ or $v_{GSR}=
-33$ km/sec (the direction and magnitude of this component of the
$v_{GSR}$ in the circular orbit case is shown as $V_{C}$ in Figure 3,
while the direction and magnitude of the {\em measured} $v_{GSR}$ for
Pyxis is shown as $V_{M}$).  Including the component of the Sun's
velocity along the line of sight ($-227$ km/sec), the heliocentric
radial velocity ($V_{C helio}$ in Figure 3) for Pyxis in this case
would be $-33 -(-227)= 194$ km/sec.  Since this is much larger than the
measured value (36 km/sec, $V_{M helio}$ in Figure 3), we conclude that
Pyxis is not in an orbit that gives rise to a present space velocity
near $\vec{V}_{circ}$ and, therefore, the strawman model of Pyxis being
on a {\em circular} orbit and sharing the LMC orbital pole grossly fails
expectations.  Thus, one or both of the assumptions in the strawman
model must be invalid:  either Pyxis is not on a circular orbit or/and
Pyxis' orbit does not share a pole with the LMC.

Since the inferred radial velocity of the Pyxis globular cluster with
respect to a stationary observer at the location of the Sun has a large
magnitude, $v_{GSR}\sim-191$ km/sec, constraints can be placed on 
non-circular orbits Pyxis may follow that also share the plane and
direction of rotation of the LMC's orbit.  We have calculated the
orbital pole for the Pyxis globular cluster given all possible,
realistic proper motions and accounting for our derived radial
velocity.  In Figure 4, we present the region in proper motion space
that produces an orbital pole for the Pyxis globular cluster that is
within $15^{\circ}$ of the pole of the LMC's orbit.  We limit the
possible proper motions to those that yield a space velocity less than
the escape velocity from the Milky Way at the position of Pyxis
($\sim\,$415 km/sec in the Galactic model of Kochanek [1996]) and obtain the
shaded region in Figure 4.  This is a prediction for the magnitude and
direction of the proper motion of Pyxis with respect to the Sun
assuming that the Milky Way capture from the Magellanic Clouds
hypothesis is correct.

The proper motions in the shaded region of Figure 4 are those that can
produce an orbital pole in the direction of $\hat{P}$ given a $v_{GSR}$
of $-191$ km/sec for Pyxis.  We have calculated the shapes and energies
of the orbits allowed for Pyxis to determine if a pole in the direction
of $\hat{P}$ is only likely for a very restricted range of conditions.
For example, is the LMC capture origin for Pyxis only viable if Pyxis
is following an extremely eccentric orbit?  In fact, a range of orbits
is possible for Pyxis given a proper motion in the shaded region in
Figure 4.  For a given elliptical orbit, the angle between the
instantaneous velocity and radius vectors varies with position along
the ellipse and that angle has a well defined minimum value for a given
orbital eccentricity.  For each space motion derived from our radial velocity
and a proper motion from Figure 4, we have determined the angle between
the velocity vector and the present Galactocentric radius vector for
Pyxis.  Assuming a closed, elliptical orbit, we can determine the lower
limit for the eccentricity of the associated orbit having the given
angle between the velocity and radius vectors at the present position
of Pyxis.  The orbits determined for Pyxis given our radial velocity
and a proper motion in the shaded region in Figure 4 have
eccentricities of $e > 0.70$, with the peak of the distribution of all
allowable eccentricities near $e \sim\,$0.8.

We note here that few of the globular clusters with measured proper
motions are following nearly circular orbits.  \cite{dd99} has compiled
all of the measured proper motions for a sample of 38 Galactic globular
clusters and integrated orbits for each cluster.  Figure 5 presents a
histogram of the orbital eccentricities that \cite{dd99} calculated for
the globular clusters in their sample.  The open histogram in Figure 5
represents the data on the whole sample, while the hatched histogram
represents the data on the 10 clusters with apoGalactica greater than
20 kpc.  Less than half of the entire sample have eccentricities of $e
< 0.5$, and, more importantly, for the outer halo globular clusters the
measured eccentricities are mostly found in the range $0.6 < e < 0.8$.
Therefore, one might conclude that it is more likely than not that
Pyxis is following an eccentric orbit and its space motion is {\em not}
perpendicular to its current position.  However, such a conclusion must
be tempered with the acknowledgement of a potential selection bias for
the latter subsample.  The majority of the globular clusters with
measured proper motions are those that are currently close to the Sun.
Therefore, the clusters with large apoGalactica that have measured
proper motions are, for the most part, currently near periGalacticon
and thus must be following eccentric orbits.  Thus, the sample of
globular clusters that make up the hatched histogram in Figure 5 may be
selected preferentially from the sample of outer halo globular clusters
on eccentric orbits.  Since few outer halo globular clusters near
apoGalacticon have measured proper motions, the true distribution of
eccentricities for outer halo globular clusters is unknown.  However,
the fact remains that a non-negligible (and perhaps dominant) fraction
of the outer halo globular cluster population is orbiting the galaxy
with eccentricities near 0.8, and since the majority of the outer halo
globular clusters with orbits integrated by \cite{dd99} have
eccentricities near $e \sim\,$0.8, it is at least conceivable that the orbit of
Pyxis has a similar eccentricity.

For our measured Pyxis radial velocity, we have integrated orbits for a
grid of $>1500$ proper motions found in the shaded region in Figure 4
in the potential of \cite{jsh95} for 10 Gyr each.  The orbital energy
of Pyxis determined from the majority of the proper motions that
produce an orbital pole at $\hat{P}$ is within the $1-\sigma$ error
bars of the orbital energy of the LMC, although the error bar is large
($E_{LMC} = -2.1\pm0.9 \times 10^{4}$ km$^{2}$/sec$^{2}$).  However,
Johnston (1998) found that in simulations of tidal stripping, debris
was found within $\pm3\Delta E$ of the parent object (see eq. 1). {\em All}
of the orbits produced from our measured radial velocity and a proper
motion in the shaded region of Figure 4 are within $\pm3\Delta E$ of
the orbital energy of the LMC.  Since the $\hat{P}$ orbits do not
require extremely unlikely constraints on the eccentricity and since
the orbital energies for these orbits are similar to expectations for
debris from the LMC, the LMC capture origin for Pyxis
remains viable.

\section{Conclusions}

It has been proposed since its discovery that the Pyxis globular
cluster may have been captured by the Milky Way from the Magellanic
Clouds.  If the space motion for Pyxis were known, a comparison of the
position of its orbital pole with respect to the LMC as well as a
comparison of its angular momentum and orbital energy to that of the
LMC would allow one to determine if the two objects share similar
orbits.  Although only one component of the space motion of Pyxis is
now measured, some constraints can be placed on its possible orbit in
the tidal capture scenario.  A circular orbit with an orbital pole at
$(l,b)=(163,-22)^{\circ}$ is completely ruled out by the measured
radial velocity.  However, we have shown here that the large radial
velocity of Pyxis with respect to a stationary observer at the position
of the Sun does not rule out the possibility that the cluster was
captured from the LMC since a reasonable range of viable orbits with $e
\sim\,$0.8 exist for Pyxis that are also similar in energy and angular
momentum to that of the LMC.  No suitable first epoch plate material is
known to exist for Pyxis, so an attempt to measure its proper motion to
better determine the likelihood that Pyxis may be a captured LMC
globular cluster will require precise observations with the HST or the
Space Interferometry Mission (SIM).

Although proper motions are not available for the majority of the outer
halo globular clusters, their spatial distribution has been used to
argue that they are likely to have been accreted into the halo (e.g.,
Majewski 1994, Palma et al.\ 2000).  Recently, Dinescu et al.\ (2000)
have measured a proper motion for the young globular cluster Pal 12 and
they find that its orbit is what one would expect if it had been
captured from the {\em Sagittarius} dwarf galaxy.  An accretion origin
of the outer halo, second parameter horizontal branch globular clusters
is often invoked to explain the possible younger age of some of these
objects (where youth is inferred either from the second parameter
effect itself or from relative age estimates determined from
the cluster CMDs).  The physical mechanism that causes the second parameter
effect in globular clusters is still unknown: Although it is now
generally agreed that there are indeed some globular clusters with
anomalously young ages, age differences alone may not be enough to
explain the second parameter effect.  Whether or not the physical
mechanism that causes the effect is age, the possibility that
conditions somehow favor the formation of second parameter globular
clusters preferentially in Milky Way satellite galaxies (which
later get accreted by the Galaxy) may explain the source of the
differences between second parameter and non-second parameter globular
clusters.

The age measurement for the Pyxis globular cluster by \cite{sg96},
$13.3\pm1.3$ Gyr, suggests that it is younger by $\sim\,$3 Gyr than the
oldest Milky Way globular clusters when measured on the same age
scale.  Recently, age measurements for the oldest LMC globular clusters
have been made (\cite{olsen98}) using a different technique than that
used for Pyxis, but their average age of $15.3\pm1.5$ Gyr places them
similar in age to the oldest Milky Way globular clusters, when
calibrated onto the same absolute age scale.  Another study of a
different sample of LMC clusters (\cite{johnson99}) also finds the
oldest LMC clusters to be as old as the old Milky Way clusters.  Thus,
we may conclude that typical LMC clusters are older than Pyxis.
However, at least one of the clusters in the Olsen et al.\ (1998)
sample is $\sim\,$2 Gyr younger than the others (NGC 1898), which makes
it similar in age to Pyxis.  Therefore, it is not impossible to place
Pyxis in the ``LMC family'' of clusters from age arguments, though it
does appear that Pyxis would be at the young end of the age range for
old LMC clusters.

It may be noted, however, that the current orbital pole of the Small
Magellanic Cloud (SMC) is also very near the intersection point of the
poles of the LMC and Pyxis (see Figure 2).  Since the SMC is more
fragile due to its weaker gravitational potential, perhaps a more
attractive origin for Pyxis is from stripping of the SMC rather than
the LMC.  Recent studies of SMC globular clusters have found that the
SMC clusters show a range in ages (e.g., \cite{sfrz98}, \cite{msf98})
including at least one cluster with an age similar to Pyxis (NGC 121).
The orbital energy of the SMC has a larger magnitude and a smaller
error bar than that of the LMC, so not all of the orbits produced from
a proper motion in the shaded region in Figure 4 have orbital energies
similar to expectations of SMC debris.  Only the orbits having proper
motions found in the inner part of the shaded region, with a total
magnitude of the proper motion of $\sim\,$0.75 mas/yr, have orbital
energies consistent with an SMC capture origin.  Since the same orbital
energy and age arguments applied to support the LMC capture origin also
apply to the SMC, we consider it a possibility that Pyxis may have been
captured from either the LMC or the SMC.

\acknowledgements 

We would like to thank Ata Sarajedini for providing us with an
electronic version of his table of CCD photometry for Pyxis stars.  We
also appreciate a helpful conversation with Knut Olsen.  We wish to
thank the anonymous referee for useful comments that improved the
manuscript.  CP and SRM acknowledge support for this research from NSF
CAREER Award grant AST-9702521, the David and Lucile Packard
Foundation, and The Research Corporation.

\appendix

\section{Definition and Sign Conventions for $V_{GSR}$}

The conversion of velocities among various reference frames is treated
in the literature and in the standard texts, such as Mihalas \& Binney
(1981).  Because we have found some confusing misuse of the standard
terminology in the literature, we provide this detailed explanation of
our sign and naming conventions.

The radial velocity that one measures for a star is the
velocity of that object with respect to the Earth.  Often, corrections
are made to this velocity to remove the motions of the Earth and Sun,
which reduces the measured radial velocity to a velocity with respect
to some standard of rest.  For example, the measured heliocentric
radial velocity ($v_{helio}$) is reduced to the radial velocity with
respect to the Local Standard of Rest ($v_{LSR}$) by removing the Sun's
peculiar velocity with respect to the LSR:

\begin{equation}
v_{LSR} = v_{helio} + [9 \cos (b) \cos (l) + 11 \cos (b) \sin (l) + 6 \sin (b)] km/sec.
\end{equation}

\noindent This velocity can be further reduced to the Galactic Standard
of Rest (GSR) by removing the Sun's orbital velocity around the Galactic
Center.  So,

\begin{equation}
v_{GSR} = v_{LSR} + [220 \cos (b) \sin (l)] km/sec.
\end{equation}

Referring to this velocity as ``$v_{GSR}$'' or a ``Galactocentric''
velocity apparently causes some confusion in the interpretation of
velocity data.  When reduced using the above two equations, the
velocity referred to as $v_{GSR}$ is the velocity of the object as seen
by a stationary observer at the position of the Sun.  The direction of
this velocity is along the line of sight between the object and the Sun
and not along the line of sight between the object and the Galactic
Center.  The latter misinterpretation of  ``Galactocentric'' velocity (as we
have found in some articles in the literature) can lead to misleading or
erroneous conclusions.

There is an additional ambiguity in the definition of $v_{GSR}$, and
that is the sign convention.  For a typical radial velocity, positive
refers to a velocity that is moving away from
the origin, and negative refers to a velocity that is approaching the
origin.  The origin for the ``Galactocentric''
radial velocity, or $v_{GSR}$, is the Sun and not the Galactic center.
Therefore, the sign convention for $v_{GSR}$ is that a positive
velocity indicates that the object is moving away from a stationary
observer at the position of the Sun and a negative velocity indicates
that the object is moving towards a stationary observer at the position
of the Sun.  The right hand sides of the two equations above are collectively
the velocity of the Sun along the line of sight.  
This sign convention introduces additional confusion
because the sign may not agree with the sign convention for the
Cartesian Galactocentric $(U,V,W)$ system, and because the sign of
the contribution of the Sun's motion, $v_{\sun LOS}$ can seem 
counterintuitive.

For example, the Pyxis globular cluster has been measured to have a
heliocentric radial velocity of $\sim\,$36 km/sec.  Using the above
equations, $v_{GSR} = -191$ km/sec for Pyxis.  The proper
interpretation of this velocity is that Pyxis is approaching a
stationary observer at the position of the Sun with a velocity
magnitude of 191 km/sec.  However, the Sun is located in Galactic
Cartesian coordinates at $(X,Y,Z) = (-8,0,0)$ and Pyxis is located at
$(X,Y,Z) =  (-13.9,-38.6,4.8)$.  Therefore, the components of this
velocity, $v_{GSR} = -191$ km/sec, in Galactic Cartesian coordinates
are positive in $X$, positive in $Y$, and negative in $Z$.  Moreover,
even though the Sun's motion is {\it increasing} the separation
of Pyxis from us (i.e., {\it increasing} the recessional velocity)
$v_{\sun LOS}$ is {\it negative}.

\clearpage

\begin{figure}

\plotone{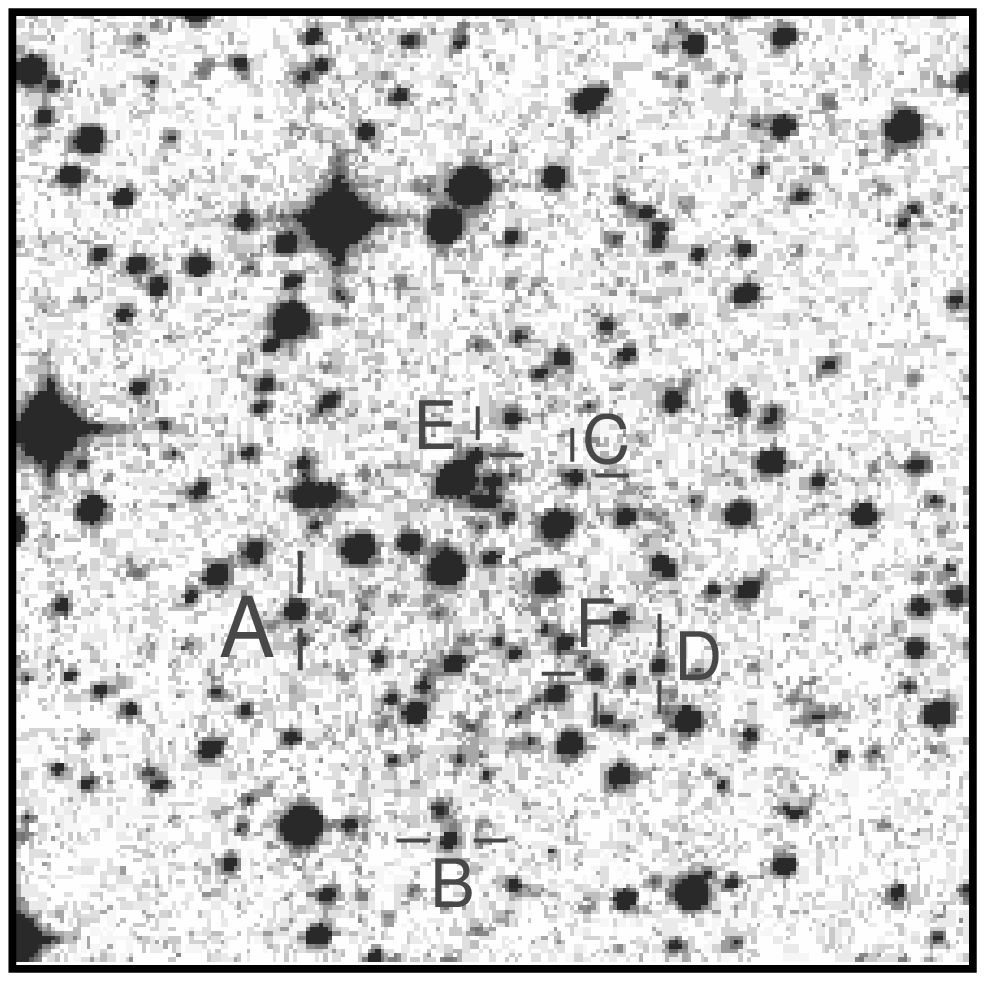}

\caption{A finding chart for the six Pyxis stars observed.  The image was created
using the Digitized Sky Survey (\cite{dss}).  In this image, the center is
$\alpha_{2000.0} = 9^{\rm h} 07^{\rm m} 57.8^{\rm s}$, 
$\delta_{2000.0} = -37\arcdeg 13\arcmin 17\arcsec$.  North is up and
East is to the left.}

\end{figure}

\clearpage

\begin{figure}

\plotone{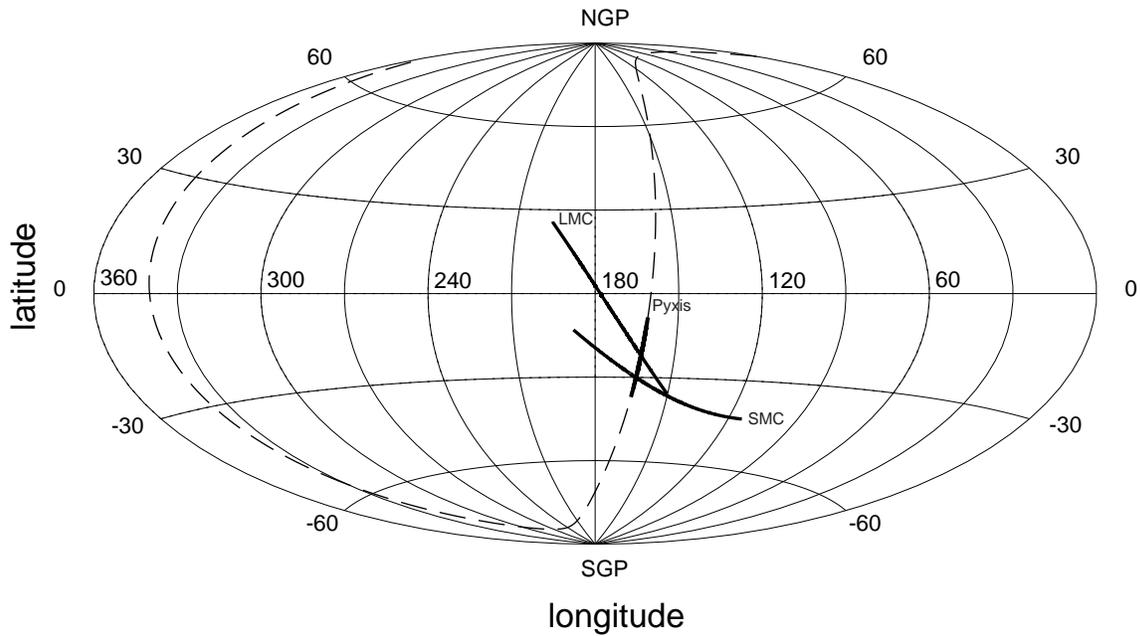}

\caption{An Aitoff projection of the sky in Galactocentric
coordinates.  The solid arc near the center traces the locations of all
possible poles for the orbit of the LMC determined from the
\cite{jones} proper motion.  The locations of the possible orbital
poles for the SMC (using the Irwin et al.\ (1996) proper motion) are
shown, as well.  The dashed great circle is the family of all possible
normals to the current location of the Pyxis globular cluster.  The
orbital pole of Pyxis is assumed to lie somewhere along this great
circle.  The solid arc along the great circle contains all points
within $15^{\circ}$ of the intersection between the LMC's possible
orbital poles and those of Pyxis.  If Pyxis has been recently captured
from the LMC by the Milky Way, its orbit should share nearly the same
pole as the LMC, and it is likely to lie along the arc shown here.}

\end{figure}

\clearpage

\begin{figure}

\plotone{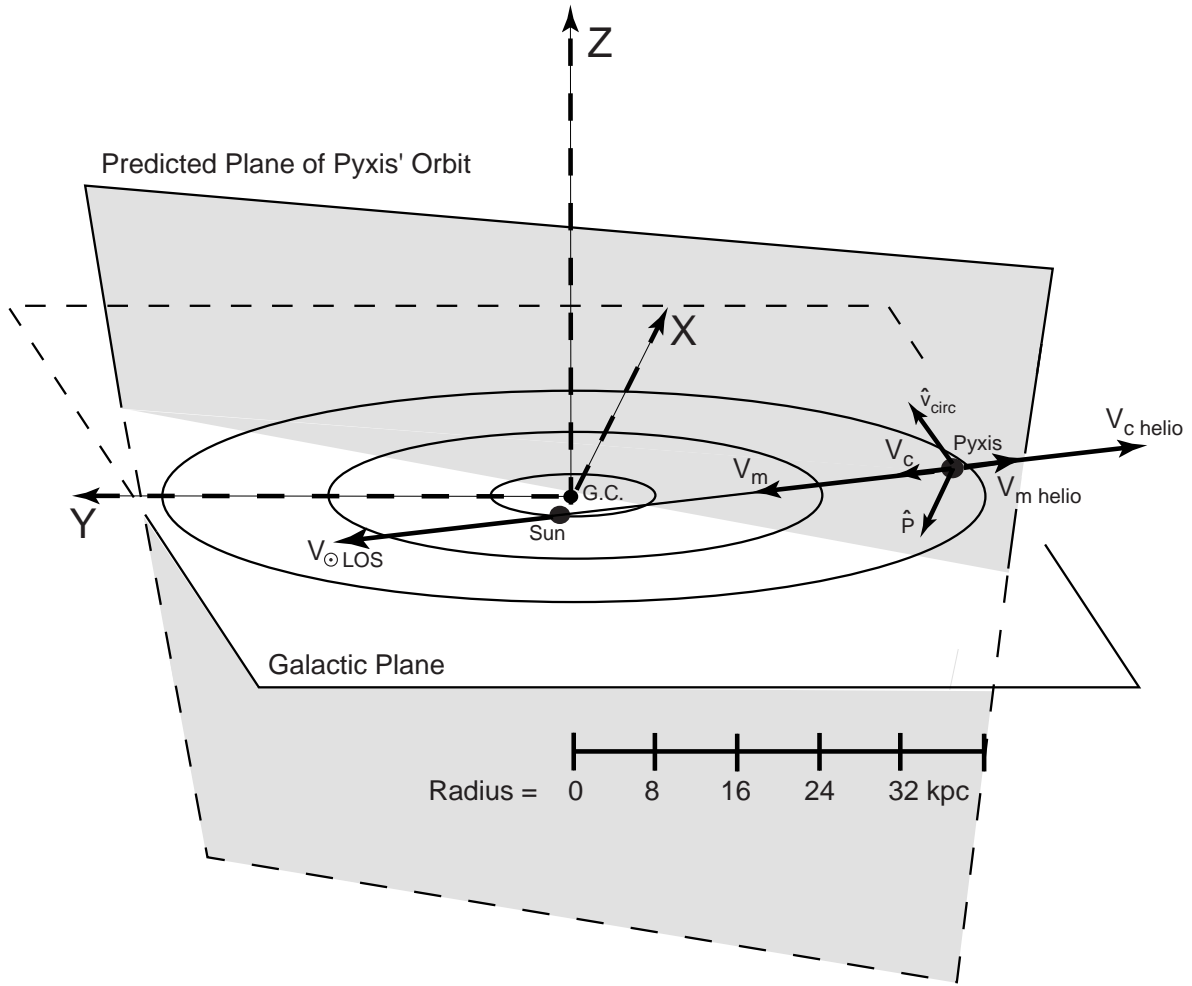}

\caption{see next page}

\end{figure}

\clearpage

\noindent Fig. 3. --- A 3-D projection of the sky in Galactic Cartesian
coordinates.  The Sun is located at $(X,Y,Z) = (-8,0,0)$ kpc and Pyxis
lies at $(-13.9,-38.6, 4.8)$ kpc. The shaded plane contains the line of
sight connecting the Galactic Center to Pyxis and is perpendicular to
the vector $\hat{P}$, which is a unit vector that points in the
direction of the orbital pole located at $(l,b) = (163,-22)^{\circ}$, a
pole that Pyxis may share with the LMC.  Any velocity vector for Pyxis
that lies in the shaded plane will give Pyxis an orbital pole in the
direction of $\hat{P}$ (by definition).   The circular motion orbital
vector $\hat{v}_{circ}$ is a unit vector that is mutually perpendicular
to both $\hat{P}$ and the Pyxis/Galactic Center line of sight and
therefore lies in the shaded plane.  If Pyxis is following a circular
orbit with its pole at $(163,-22)^{\circ}$, then its space velocity
should lie along $\hat{v}_{circ}$.  Assuming that the space velocity of
Pyxis has a magnitude near the circular velocity of the Galaxy at 40
kpc, or $\sim\,$200 km/sec, the predicted radial velocity (as seen by a
stationary observer at the location of the Sun) for Pyxis is $V_{c}$,
or $-33$ km/sec, which corresponds to a heliocentric radial velocity
($V_{c\ helio}$) of 194 km/sec.  Given that the solar motion along the
line of sight to Pyxis ($V_{Sun}$) is $-227$ km/sec and that the
measured heliocentric radial velocity of Pyxis ($V_{m\ helio}$) is 36
km/sec, the measured radial velocity of Pyxis with respect to a
stationary observer at the location of the Sun ($V_{m}$), is $-191$
km/sec, ruling out a {\em circular} orbit for Pyxis with a pole in the
direction of $\hat{P}$.  The unknown proper motion of Pyxis is only
constrained to be perpendicular to $V_{m}$ (by definition).  It is
plausible that Pyxis may be following an eccentric orbit, as is the
case for the majority of the Galactic globular cluster population, in
which case the space velocity of Pyxis is unlikely to be perpendicular
to its radius vector.  There exists a set of proper motions for Pyxis
(see Figure 4) that, when combined with $V_{m}$, produce a space motion
with a pole at $(163,-22)^{\circ}$ and leave the cluster bound to the
Milky Way.

\clearpage

\begin{figure}

\plotone{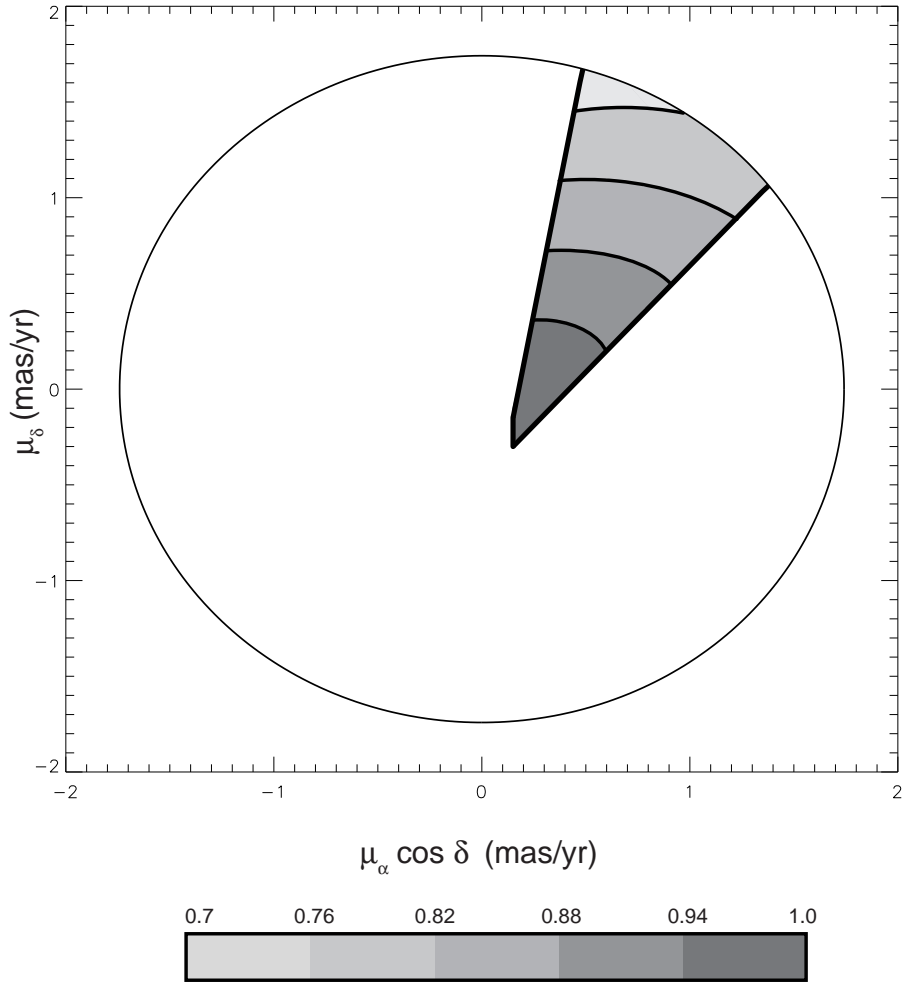}

\caption{A plot of possible proper motions in equatorial coordinates
for the Pyxis globular cluster.  The interior of the circle defines the
region of values for the transverse velocity that gives Pyxis a total
space velocity less than the escape velocity from the Galaxy ($\sim\,$415
km/sec).  The shaded region defines the permissible proper motions that
result in Pyxis having an orbital pole within $15^{\circ}$ of the
orbital pole of the LMC.  If Pyxis has been recently captured by the
Milky Way, its proper motion should lie in this region.  Given a proper
motion in the shaded region and our measured radial velocity, the
minimum eccentricity of the possible orbits will vary from $\sim0.70 -
1.0$.  The greyscale indicates the range of minimum eccentricities of
the orbits given a proper motion in the various shaded areas.  The
scale is given by the bar below the plot.}

\end{figure}

\clearpage

\begin{figure}

\plotfiddle{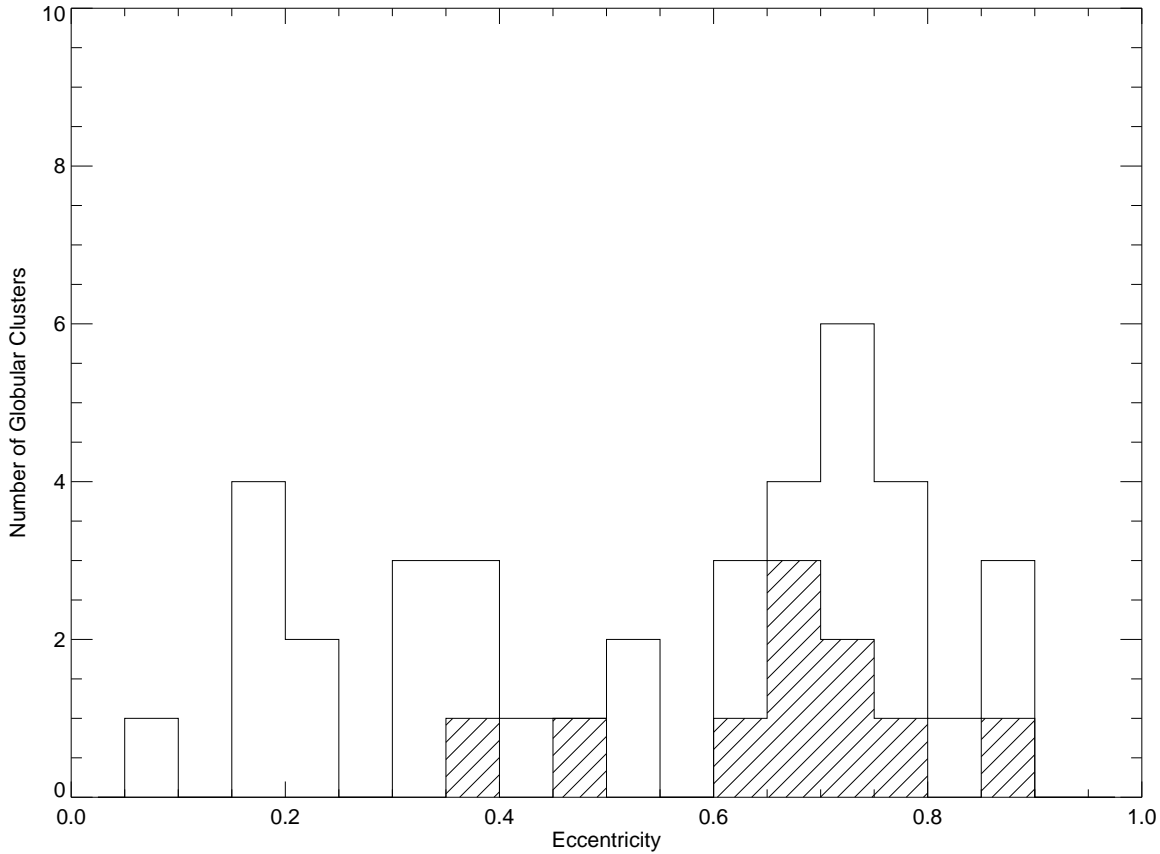}{3.5truein}{90}{67}{67}{250}{0}

\caption{The distribution of orbital eccentricities for globular clusters
in the sample of \cite{dd99}.  The open histogram represents the eccentricities
for the entire sample of 38 globular clusters with known orbits.  The hatched
histogram represents the distribution of eccentricities for the 10 objects in
the sample that have $R_{apo} > 20$ kpc, i.e., the outer halo objects.  Given
our large measured radial velocity for Pyxis, the only allowed orbits it may
follow that nearly share the orbital pole of the LMC have $e \sim\,$0.8.  Since
the majority of the outer halo clusters in the sample of \cite{dd99} have 
$0.6 < e < 0.8$, it is plausible to assume that Pyxis may also follow such an
eccentric orbit.}

\end{figure}

\clearpage

\begin{deluxetable}{lllcccc}
\tablenum{1}
\tablecaption{Observed Pyxis Giant Stars}
\tablehead{ \colhead{ID} & \colhead{$\alpha_{2000.0}$}  & \colhead{$\delta_{2000.0}$} 
& \colhead{\# of Obs.} & \colhead{$v_{helio}$}
& $R$\tablenotemark{a} & $(B-R)$\tablenotemark{a} \nl
& \colhead{(h:m:s)} & \colhead{(d:m:s)} &  & \colhead{(km/sec)} & &}
\startdata
Pyxis A &  09:08:01.7 & -37:13:48 & 4  & 32.7 & 17.08 & 2.01 \nl
Pyxis B &  09:07:58.6 & -37:14:42 & 2  & 38.4 & 17.75 & 1.80 \nl
Pyxis C &  09:07:56.2 & -37:13:16 & 4  & 36.6 & 18.28 & 1.78 \nl
Pyxis D &  09:07:54.5 & -37:14:01 & 5  & 26.1 & 18.33 & 1.75 \nl
Pyxis E &  09:07:57.8 & -37:13:17 & 4  & 37.9 & 18.08 & 1.78 \nl
Pyxis F &  09:07:55.7 & -37:14:02 & 3  & 33.9 & 18.09 & 1.71 \nl
\enddata

\tablenotetext{a}{Photometry from \cite{sg96}.}

\end{deluxetable}

\end{document}